\documentclass{PoS}

\usepackage{graphics}{}
\usepackage[pdftex]{}
\usepackage{amsmath,amssymb,amsfonts}

\usepackage{epsfig}
\usepackage{amssymb}
\usepackage{subfigure}

\newcommand{\agev}{\mbox{$A$~GeV}}               
\newcommand{\gevc}{\mbox{GeV$/c$}}

\newcommand{\mevc}{\mbox{MeV$/c$}}

\newcommand{\rb}[1]{\mbox{\textrm{\scriptsize #1}}}

\newcommand{\pt}{\ensuremath{p_{\rb{T}}}}

\title{NA61/SHINE low energy program at SPS}

\ShortTitle{NA61/SHINE low energy program at SPS}

\author{\speaker{Grzegorz~Stefanek} for the NA61 Collaboration\\
        Jan Kochanowski University, Kielce, Poland\\
        E-mail: \email{grzegorz.stefanek@pu.kielce.pl}

\vspace{0.5cm} \noindent
N.~Abgrall${}^{22}$,
A.~Aduszkiewicz${}^{23}$,
B.~Andrieu${}^{11}$,
T.~Anticic${}^{13}$,
N.~Antoniou${}^{18}$,
J.~Argyriades${}^{22}$,
A.~G.~Asryan${}^{15}$,
B.~Baatar${}^{9}$,
A.~Blondel${}^{22}$,
J.~Blumer${}^{5}$
% ...
}

\author{\\
\noindent
${}^{ 1}$Cape Town University, Cape Town, South Africa \\
${}^{ 2}$ETH, Zurich, Switzerland \\
${}^{ 3}$Fachhochschule Frankfurt, Frankfurt, Germany \\
${}^{ 4}$Faculty of Physics, University of Sofia, Sofia, Bulgaria \\
${}^{ 5}$Forschungszentrum Karlsruhe, Karlsruhe, Germany \\
${}^{ 6}$Institute for Nuclear Research, Moscow, Russia \\
${}^{ 7}$Institute for Particle and Nuclear Studies, KEK, Tsukuba,  Japan \\
${}^{ 8}$Jagiellonian University, Cracow, Poland  \\
${}^{ 9}$Joint Institute for Nuclear Research, Dubna, Russia \\
${}^{10}$KFKI Research Institute for Particle and Nuclear Physics, Budapest, Hungary \\
${}^{11}$LPNHE, University of Paris VI and VII, Paris, France \\
${}^{12}$University of Silesia, Katowice, Poland  \\
${}^{13}$Rudjer Boskovic Institute, Zagreb, Croatia \\
${}^{14}$Soltan Institute for Nuclear Studies, Warsaw, Poland \\
${}^{15}$St. Petersburg State University, St. Petersburg, Russia \\
${}^{16}$State University of New York, Stony Brook, USA \\
${}^{17}$Jan Kochanowski University in  Kielce, Poland \\
${}^{18}$University of Athens, Athens, Greece \\
${}^{19}$University of Bergen, Bergen, Norway \\
${}^{20}$University of Bern, Bern, Switzerland \\
${}^{21}$University of Frankfurt, Frankfurt, Germany \\
${}^{22}$University of Geneva, Geneva, Switzerland \\
${}^{23}$University of Warsaw, Warsaw, Poland \\
${}^{24}$Warsaw University of Technology, Warsaw, Poland  \\
}

\abstract{Status of the new experimental program to study hadron production in hadron-nucleus and nucleus-nucleus collisions at the CERN SPS will be presented.
A physics motivation will be given for the part of the program related to the physics of strongly interacting matter. Upgrades of the NA61 experimental set-up and data taking plans will be described in detail.}

\FullConference{5th International Workshop on Critical Point and Onset of Deconfinement - CPOD 2009,\\
         June 08 - 12 2009\\
         Brookhaven National Laboratory, Long Island, New York, USA}

\begin{document}

\section{Introduction}
 The NA61/SHINE \cite{NA61_webpage} (SHINE = SPS Heavy Ion and Neutrino Experiment) is a new fixed-target experiment at the CERN SPS accelerator. The most important and expensive components of the NA61 detector are inherited from the NA49 experiment although several significant upgrades have been or will be introduced.The experiment was approved at CERN in June 2007 and the first pilot run was performed in October 2007. The history of the experiment is documented in \cite{Antoniou:2006qz,Antoniou:2006mh,Antoniou:2007ad1,Antoniou:2007zzd,Abgrall:2007zza,
 Abgrall:2009sec}.

\section{Physics goals}

 The physics programme of NA61 is the systematic measurement of hadron production in proton-proton, proton-nucleus and nucleus-nucleus collisions. The programme consists of three subjects. In the first stage of data taking (2007-2009) measurements are performed of hadron production in proton-nucleus interactions needed for neutrino and cosmic ray experiments. In the second stage (2009-2010) hadron production in proton-proton and proton-nucleus interactions as reference data for better understanding of nucleus-nucleus reactions will be studied.

\begin{figure}[!ht]
\begin{center}
\begin{minipage}[b]{1.\linewidth}
\begin{center}
{\resizebox{10 cm}{!}{
\includegraphics{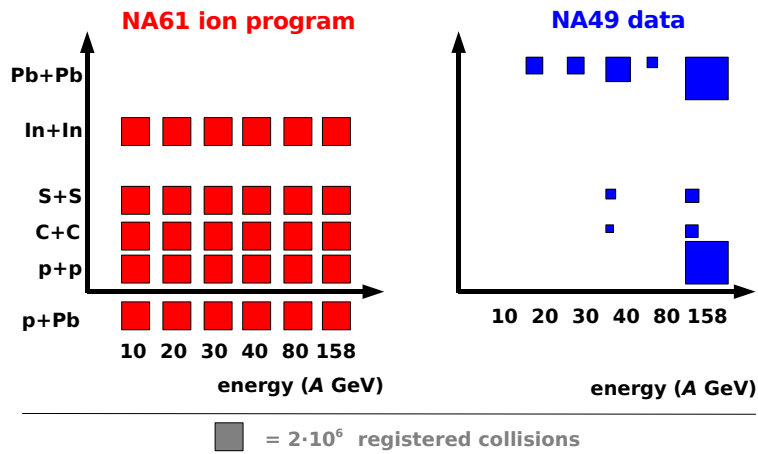}
} }
\end{center}
\end{minipage}
\end{center}
\caption{Data sets planned to be recorded by NA61/SHINE (left) within the ion program and those recorded by NA49 (right). The area of the boxes is proportional to the number of registered central collisions, which for NA61/SHINE will be $2 \cdot 10^6$ per reaction and energy.}
\label{runs}
\end{figure}

 In the third stage (2009-2013) the energy dependence of hadron production properties will be measured in nucleus-nucleus collisions as well as p+p and p+Pb interactions. The aim is to identify the properties of the onset of deconfinement and find the critical point of strongly interacting matter. The last physics goal requires a comprehensive scan in the whole SPS energy range from 10A to 158\agev~ with light and intermediate mass nuclei. NA61/SHINE intends to register p+p, C+C, S+S, In+In and p+Pb collisions at six energies with a typical number of recorded central collision events per reaction and energy of $2\times10^{6}$. The data sets planned to be recorded by NA61/SHINE for the ion program and those recorded by NA49 are compared in Fig.~\ref{runs}.

\subsection{Study of the properties of the onset of deconfinement}

\begin{figure}[!ht]
\begin{center}
\begin{minipage}[b]{1.\linewidth}
\begin{center}
\vspace*{-10.95cm}
{\resizebox{13 cm}{!}{
\includegraphics[width=0.8\textwidth]{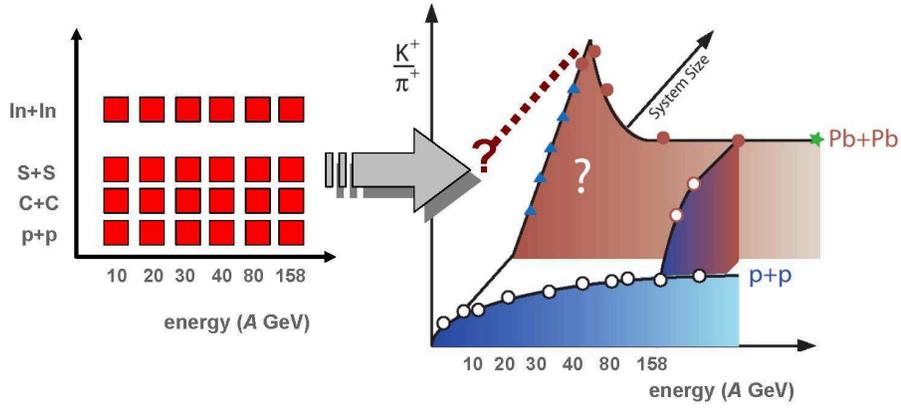}
} }
\end{center}
\end{minipage}
\end{center}
\vspace*{-0.5cm}
\caption{An illustration of the impact of the new measurements (of central collisions) on clarifying the system size dependence of the $K^{+}/\pi^{+}$ horn observed in central Pb+Pb collisions at low SPS energies.}
\label{na61-onset}
\end{figure}

 Recent results on the energy dependence of hadron production in central Pb+Pb collisions at 20, 30, 40, 80 and 158\agev~coming from the energy scan program at the CERN SPS provide evidence for the onset of a transition to the deconfined Quark Gluon Plasma phase. Hadron production properties in central Pb+Pb (Au+Au) collisions as a function of collision energy exhibit rapid changes at about 30\agev~ \cite{Afanasiev:2002mx,:2007fe}. These anomalies were predicted for the onset of deconfinement \cite{Gazdzicki:1998vd} and their further understanding requires new NA61 data. An illustration of the impact of the new measurements on the system size dependence of the $K^{+}/\pi^{+}$ horn is shown in Fig.~\ref{na61-onset}.

\subsection{Search for the critical point of strongly interacting matter}
 Lattice QCD calculations \cite{fodorkatz} indicate that the phase diagram of strongly interacting matter features a first order phase transition boundary in the temperature ($T$) - baryochemical ($\mu_{B}$) potential plane, which has a critical endpoint. This critical endpoint may be located in the energy range accessible at the CERN SPS.

 The temperature and baryochemical potential are not directly measurable quantities but the $T - \mu_{B}$ coordinates of the freeze-out points of nuclear reactions can be brought into one-to-one correspondence with the energy ($E$) and system size ($A$) of nuclear collisions \cite{Becattini:2005xt}. Therefore, the $T - \mu_{B}$ coordinates of the freeze-out points may be scanned via a systematic $E - A$ scan. When the freeze-out point of a given reaction is near the critical point, the increase of multiplicity and transverse momentum fluctuations is expected \cite{Stephanov:1999zu}. The scaled variance $\omega$ of the multiplicity distribution is expected to increase by more than 0.1 and the $\Phi_{P_{T}}$ measure of transverse momentum fluctuations by about 10~\mevc~in the standard NA49 acceptance \cite{Stephanov:1999zu,Anticic:2003fd}. An illustration of the impact of the NA61 measurements of central collisions on the search for the critical point of strongly interacting matter is shown in Fig.~3.

\begin{figure} [htb]
\begin{center}
\vspace*{-10.95cm}
\includegraphics[width=0.8\textwidth]{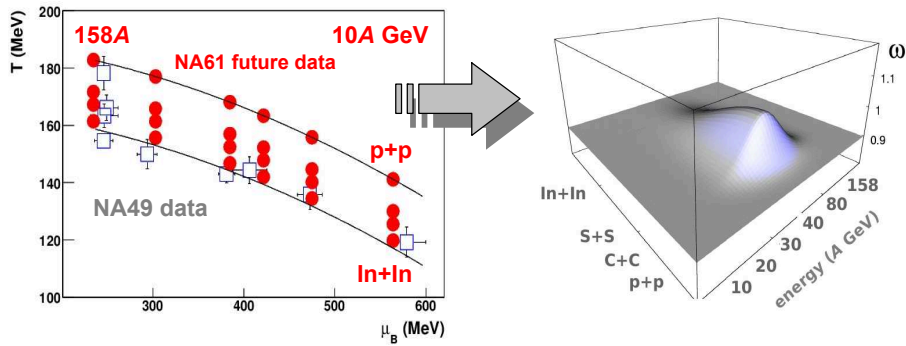}
\vspace*{-0.3cm}
\caption{An illustration of the impact of the new measurements of central collisions (the red points in the plane $T,\mu_{B}$) on the search for the critical point of strongly interacting matter.}
\end{center}
\label{na61-cp}
\end{figure}

 Measurements of central C+C, Si+Si collisions and p+p interactions at the top SPS energy as well as central Pb+Pb interactions at 5 different energies were done by the NA49 experiment \cite{Grebieszkow:2009jr}. The energy dependence of $\Phi_{P_{T}}$ and $\omega$ (see Fig.~4) show no indication for critical point fluctuations, however a narrower $\mu_{B}$ scan would be desirable.

\begin{figure}[ht]
\centering
\vspace{-0.35cm}
\subfigure[Energy dependence of $\Phi_{p_T}$ for the 7.2\% most central Pb+Pb collisions in the forward-rapidity region $1.1 < y^{*}_{\pi} < 2.6$ and $0.005 < p_T <
1.5$ GeV/c; $y^{*}_p < y^{*}_{beam} -0.5$ (to reject the projectile spectator domain) and common azimuthal angle acceptance. Lines correspond to $CP_1$ predictions ($T\approx$ 147 MeV, $\mu_{B}\approx$ 360 MeV) added to the energy averaged $\Phi_{p_T}$ measurement.]{
\includegraphics[scale=0.6]{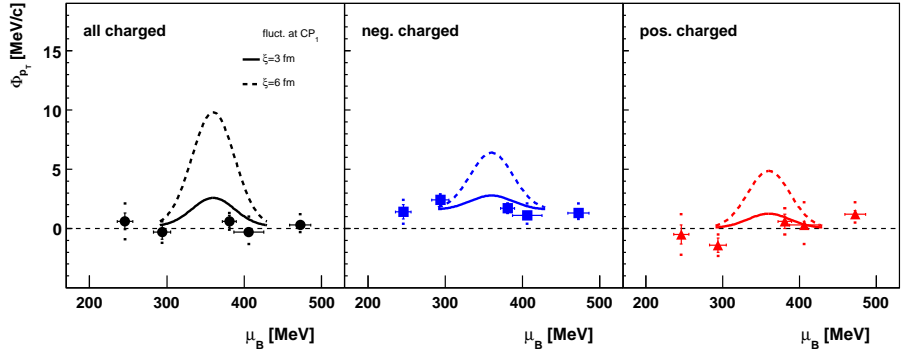}
\label{fiptmb}
}
\vspace{-0.45cm}
\subfigure[Energy dependence of $\omega$ for the 1\% most central Pb+Pb collisions in the forward-rapidity region $1.1 < y^{*}_{\pi} < y_{beam}$ and NA49 azimuthal angle acceptance. Lines correspond to $CP_1$ predictions ($T\approx$ 147 MeV, $\mu_{B}\approx$ 360 MeV) added to the energy averaged $\omega$ measurement.]{
\includegraphics[scale=0.6]{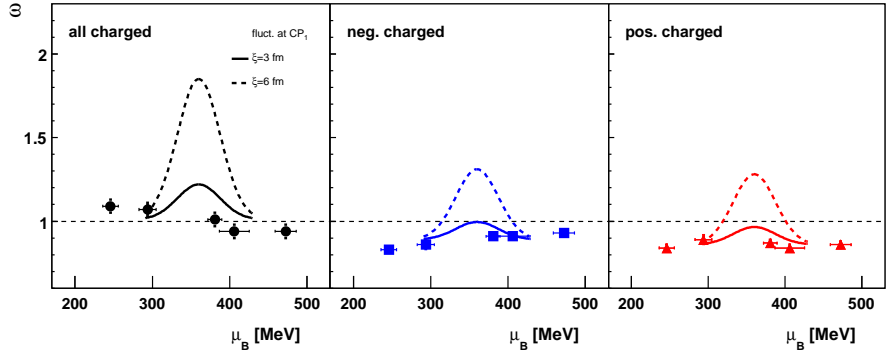}
\label{omegamb}
}
\vspace{0.5cm}
\label{fig:mb}
\caption[]{Energy dependence of fluctuations for central Pb+Pb collisions registered by NA49.}
\end{figure}

 A maximum of mean \pt~ and multiplicity fluctuations as a function of the system size is observed at 158\agev~ for smaller systems Si+Si, C+C (see Fig.~5). If one takes the results as an indication of the critical point it would be located at $T\approx$ 178 MeV and $\mu_{B}\approx$ 250 MeV \cite{Grebieszkow:2009jr}. A detailed NA61 energy and system size scan is necessary to establish the existence of the critical point.

\begin{figure}[ht]
\centering
\vspace{-0.35cm}
\subfigure[System size dependence of $\Phi_{p_T}$ at 158$A$ GeV showing results from p+p, semi-central C+C (15.3\%) and Si+Si (12.2\%), and 5\% most central Pb+Pb collisions. Forward-rapidity region $1.1 < y^{*}_{\pi} < 2.6$ and $0.005 < p_T < 1.5$ GeV/c; NA49 azimuthal angle acceptance. Lines correspond to $CP_2$ predictions ($T\approx$ 178 MeV, $\mu_{B}\approx$ 250 MeV) shifted to reproduce the $\Phi_{p_T}$ value for central Pb+Pb collisions.]{
\includegraphics[scale=0.6]{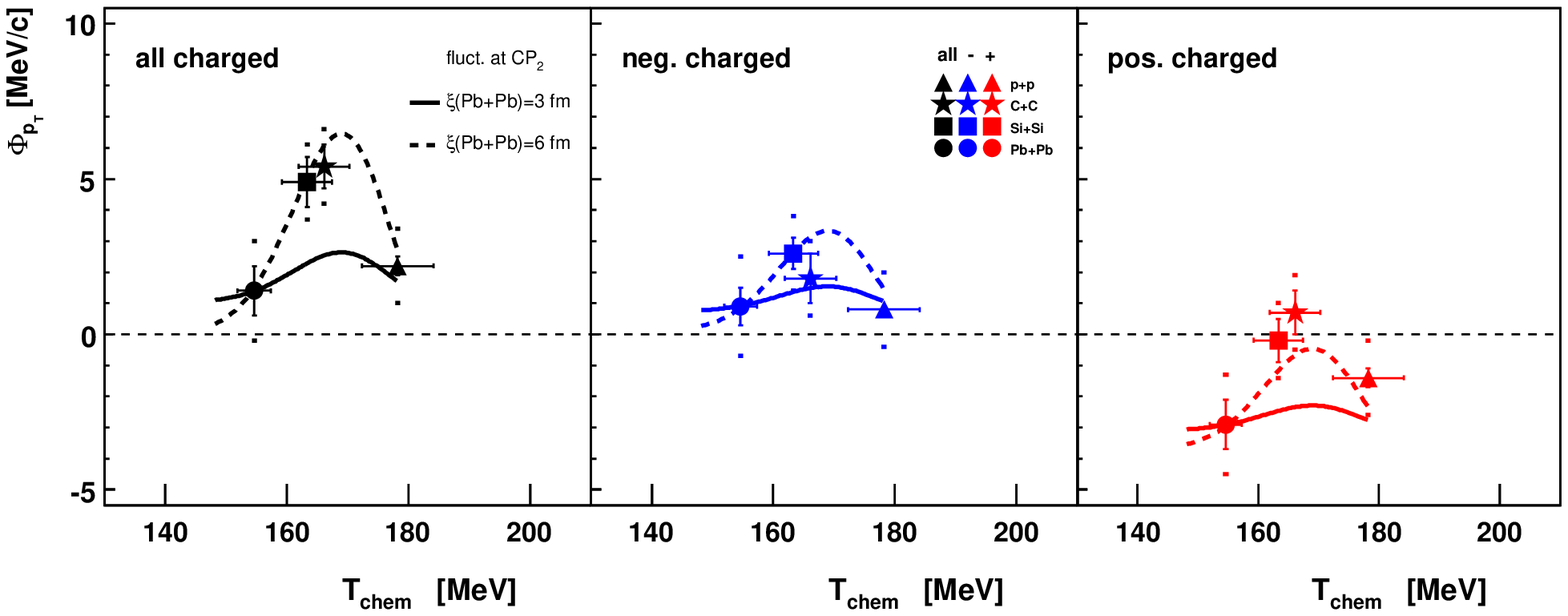}
\label{fiptT}
}
\vspace{-0.45cm}
\subfigure[System size dependence of $\omega$ at 158$A$ GeV for the 1\% most central p+p , C+C and Si+Si , and Pb+Pb collisions. Forward-rapidity region $1.1 < y^{*}_{\pi} < y_{beam}$ ($1.1 < y^{*}_{\pi} < 2.6$ for p+p); NA49 azimuthal angle acceptance. Lines correspond to $CP_2$ predictions ($T\approx$ 178 MeV, $\mu_{B}\approx$ 250 MeV) shifted to reproduce the $\omega$ value for central Pb+Pb collisions.]{
\includegraphics[scale=0.6]{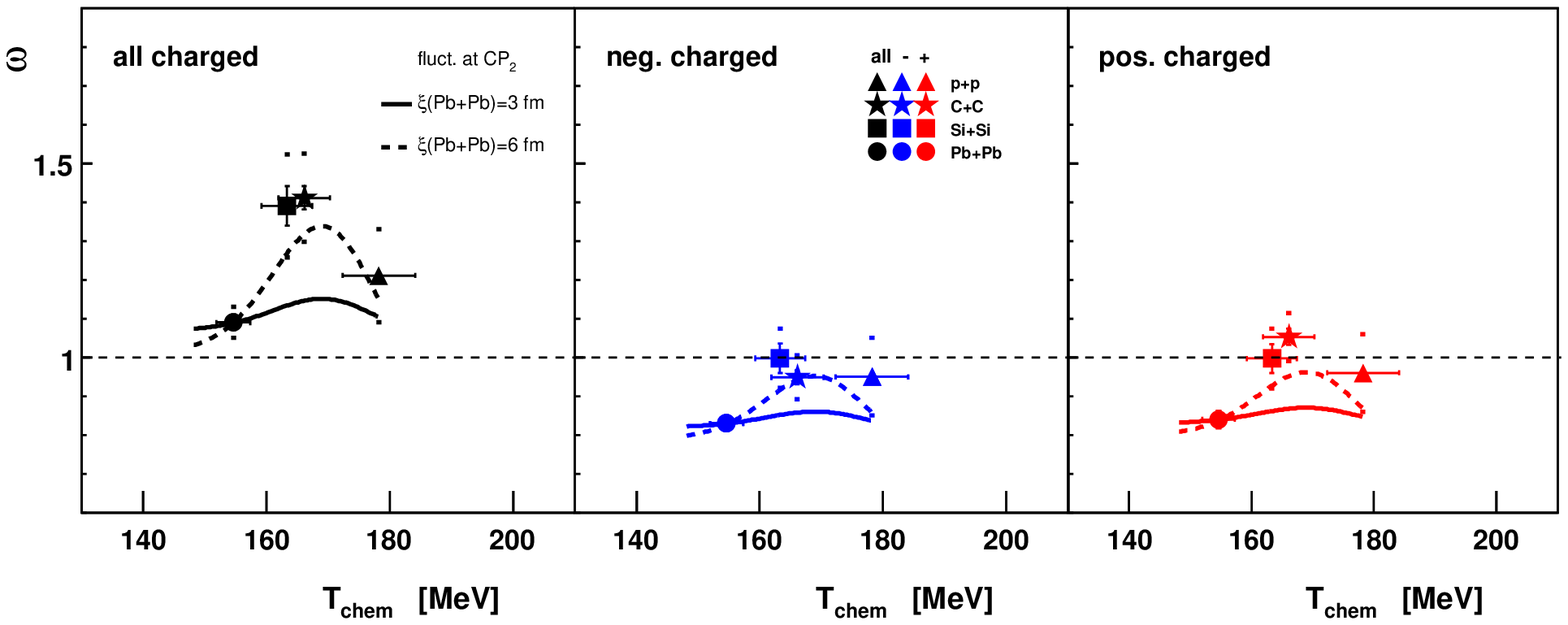}
\label{omegaT}
}
\vspace{0.5cm}
\label{fig:T}
\caption[]{System size dependence of fluctuations for central Pb+Pb collisions registered by NA49.}
\end{figure}

\section{The NA61 detector}

\begin{figure}
\begin{center}
\vspace*{-9.95cm}
\includegraphics[width=0.8\textwidth]{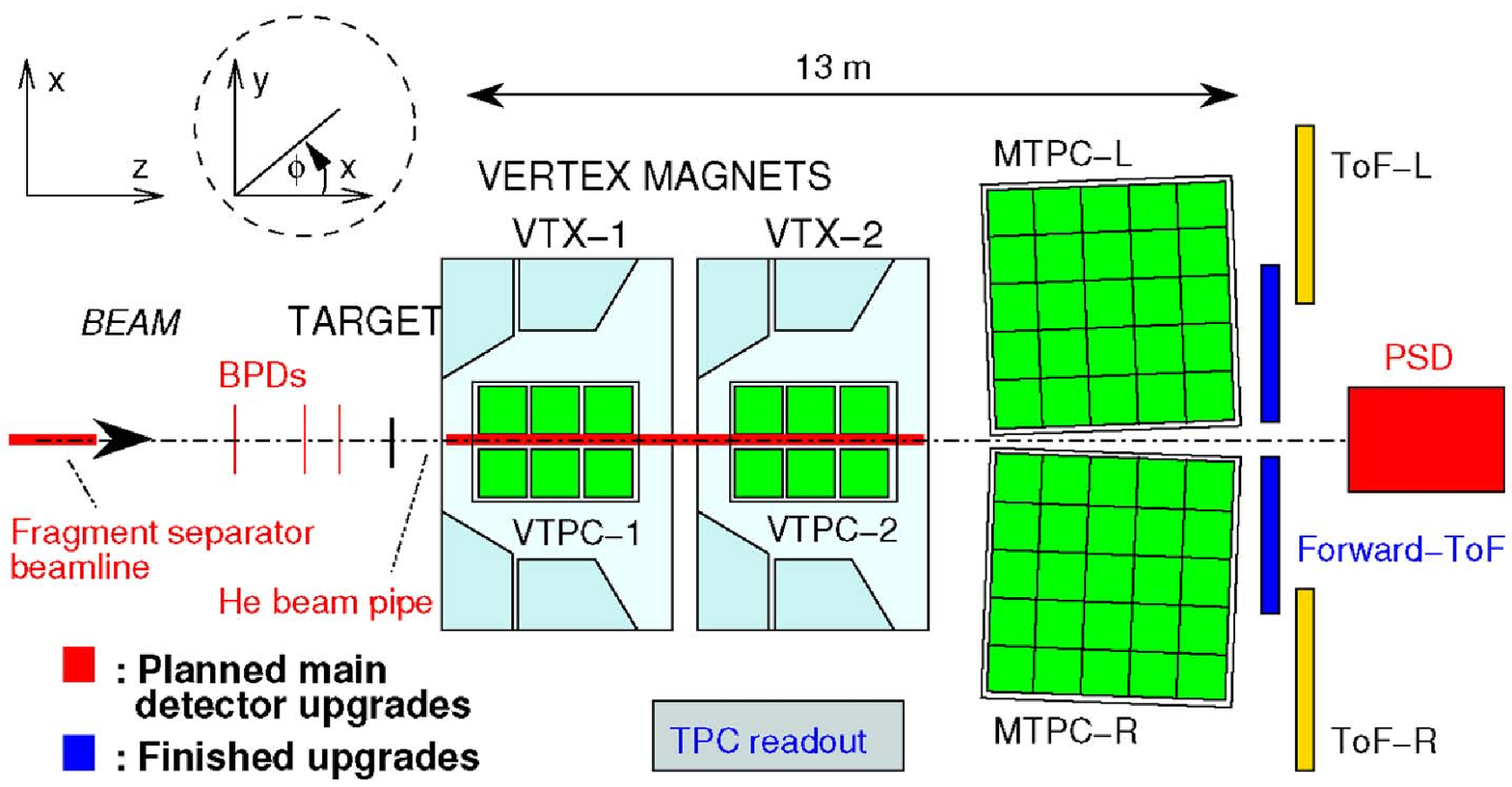}
\vspace*{-0.1cm}
\caption{The layout of the NA61/SHINE set-up (top view, not to scale) with the basic future upgrades indicated in red and finished upgrades indicated in blue.}
\end{center}
\label{na61-setup}
\end{figure}

 The NA61 experimental setup is shown in Fig.~6. The main components of the NA61 detector are four large-volume Time Projection Chambers for tracking and particle identification. The TPC system consists of two vertex chambers inside the spectrometer magnets which allow separation of positively and negatively charged tracks and a precise measurement of the particle momenta. Two main chambers, placed behind the magnets at both sides of the beam, were optimized for high precision detection of the ionization loss $dE$/$dx$ with a resolution of 3$-$6\%.

\subsection{Performance of the detector}

 During the 2007 pilot run no significant deterioration of the excellent detection capabilities of the NA61/SHINE apparatus was observed. The most important performance parameters of the detector are:
\begin{itemize}
\item[-] Large acceptance $\approx50\%$ at $\pt \lesssim 2.5~\gevc$,
\item[-] Precise momentum measurement
$\sigma(p)/p^{2}$ = $(0.3-7)\times10^{-4}(\gevc)^{-1}$,
\item[-] High tracking efficiency > 95\%,
\item[-] Good particle identification ToF resolution
$\sigma(t)\approx$ 60 ps, $dE$/$dx$ resolution $\sigma$($dE$/$dx$) $\approx5\%$, invariant mass resolution $\sigma(m)\approx$ 5 MeV.
\end{itemize}

\subsection{Main upgrades}
 The important NA61/SHINE detector upgrades are motivated by the physics goals. For the 2007 pilot run a new forward time-of-flight detector (ToF-F) was constructed in order to extend the acceptance of the NA61 set-up for pion and kaon identification as required for T2K measurements. Furthermore, numerous small modifications and upgrades of the NA61/SHINE facility were performed before the 2007 run: speed-up of the ToF-L/R readout, modification of the DAQ system to allow writing data on disk, refurbishment of the Beam Position Detectors (BPD-1/2/3) and installation of new beam counters for a new trigger logic. The new TPC readout and DAQ system with about 70 Hz readout frequency were installed and tested during a 2008 test run. These modifications increased the event rate by a factor of about 10.

  One of the future upgrades planed for the ion runs is the construction of the Projectile Spectator Detector (PSD) \cite{Golubeva:2009zz}. Compared to the NA49 veto calorimeter it provides an increase of the resolution of the measurement of the number of projectile spectators by a factor of 5 to about $dE$/$dx$ $\approx$ 50$\%$/$E$. Tests of a PSD supermodule, i.e. a $3\times3$ module matrix, were performed during 2007 and 2008. The assembling of the whole PSD is planned for 2011.

 Channeling of the high intensity heavy ion beam through the gas volume of the Vertex TPCs has limitations when compared to the proton beam. Delta electrons produced in the gas volume inside the VTPCs from heavy ion beam-gas interactions may significantly increase the background in the TPCs and distort measurements of event-by-event fluctuations. A new low mass helium beam pipe will reduce the delta electron background by a factor of 10.

\begin{figure}
\begin{center}
\vspace{-0.45cm}
\includegraphics[width=0.8\textwidth]{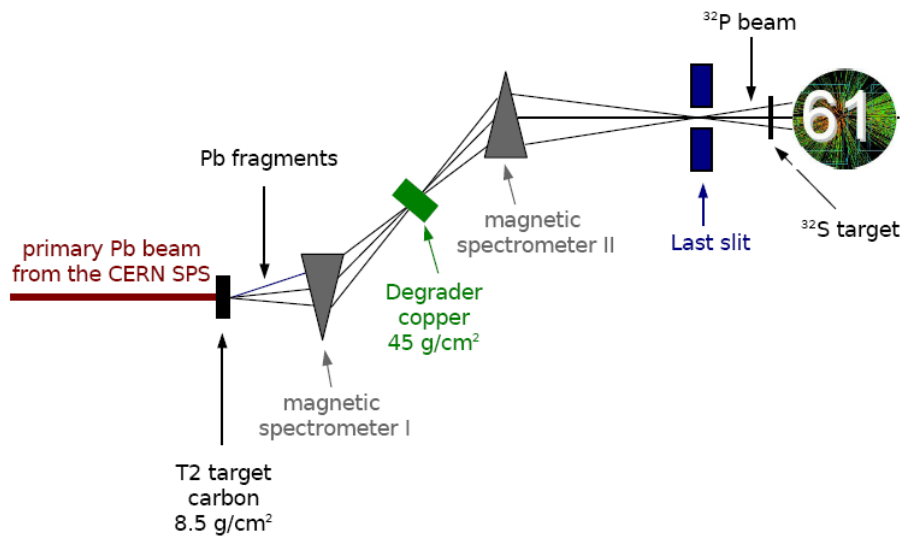}
\caption{Schematics of the proposed fragment separator in the H2 beam line (side view, not to scale). As an example selected trajectories of $^{32}P$ fragments are indicated by thin solid lines.}
\end{center}
\label{beamline}
\vspace{-0.45cm}
\end{figure}

 The NA61/SHINE physics program requires low and intermediate mass ion beams. However, only Pb beams may be available in some years. Thus the use of a secondary ion beam derived from the fragmentation products of primary Pb ions is proposed \cite{Abgrall:2009sec}. The main and essential components of the proposed beam line are the first vertical bending section which acts as rigidity filter, the degrader at the first focus, the second bending section which separates different charge states due to Z-dependent energy loss in the degrader and the last slit which selects the wanted ions (see Fig.~7). Detailed simulation of the secondary ion beam line proved that ions with $Z$=15 selected by the NA61 trigger consist mainly (75\%) of $A$=32 ions \cite{Abgrall:2009sec}.

\section{Status and plans}

 A pilot run in 2007 and the second run performed in 2008 showed that the detector fulfills the physics requirements and the finished upgrades significantly extend the detection capabilities of the NA49/SHINE setup.
 The NA61/SHINE data taking plan is presented in Table \ref{beam2}, together with the current recommendation and approval status assigned by the SPS Committee and the CERN Research Board. Three runs with ion beams are planned with nuclear mass number of $A$ $\approx$ 10, 30, 100. The sequence of data taking is optimized to increase the probability to observe indications of the new physics in the shortest time. From this point of view the most promising strategy is to start ion data taking in 2011 with $A\approx30$ and continue in 2012 and 2013 with lighter and heavier ions.

 \begin{table}[!ht]
\begin{center}
\begin{tabular}
{ c c c r r r r r }
\hline
  Beam &  Beam     &Target &  Energy    &  Year  & Days &  Physics & Status\\
  Primary  & Secondary &    & ($A$ GeV)  &        &      &       &   \\
\hline
\hline
 p &  & & 400 & &  \\
  & p & C(T2K)  &  31   &  2009  &  21
&   T2K, C-R & \it recommended  \\
\hline
 p &  & & 400 & &  \\
   & $\pi^-$ & C &  158,350   &  2009  &  2x7
&   C-R & \it recommended \\
\hline
 p &  & & 400 & &  \\
  & p & p  &  10,20,30,40,80,158  &  2009  & 6x7
&  CP\&OD & \it recommended   \\
\hline
\hline
 p &  & & 400 & &  \\
   & p & p &  158          &  2010  &  77
&  High p$_T$ & \it recommended   \\
\hline
\hline
 Pb & & & 10,20,30,40,80,158 & &  \\
 & $A\approx30$ & $A\approx30$  &  10,20,30,40,80,158  &  2011  & 6x7
&  CP\&OD  & \it recommended  \\
\hline
 p &  & & 400 & &  \\
   & p & Pb  &  158          &  2011  &  6x7
&  High p$_T$ & \it recommended   \\
\hline
\hline
 Pb & & & 10,20,30,40,80,158 & &  \\
    & $A\approx10$ & $A\approx10$   &  10,20,30,40,80,158
&  2012  &   6x7
&  CP\&OD  & \it to be discussed \\
\hline
 p &  & & 400 & &  \\
   & p & Pb  &  10,20,30,40,80,158  &  2012  &  6x7
&   CP\&OD  & \it recommended \\
\hline
\hline
 Pb & & & 10,20,30,40,80,158 & &  \\
    & $A\approx100$ & $A\approx100$  &  10,20,30,40,80,158
& 2013  &   6x7
&  CP\&OD  & \it to be discussed  \\
\hline
\hline
 & & & &  \\
\end{tabular}
\end{center}
\caption[dummy]{
The NA61/SHINE  data taking plan.
The runs with ion beams are planned for 2011, 2012 and
2013. In these runs the nuclear mass number of the selected ions
will be $A \approx 30$,  $A \approx 10$ and  $A \approx 100$, respectively.
The following abbreviations are used for the
physics goals of the data taking: CP - Critical Point, OD - Onset of
Deconfinement, C-R - Cosmic Rays.
}
\label{beam2}
\end{table}

\section{Summary}
The NA61/SHINE experiment will first perform measurements of hadron production in hadron-nucleus interactions needed for neutrino and cosmic ray experiments. The main part of the program will study energy and nuclear mass dependence of hadron production in nucleus-nucleus collisions with the aim to identify properties of the onset of deconfinement. It has a significant discovery potential for the critical point of strongly interacting matter, if it exists. There are also several future projects developed at BNL, FAIR and NICA complementary to NA61 which will provide experimental data for study of strongly interacting matter in the region of the onset of deconfinement.

\begin{acknowledgments}
This work was supported by the Polish Ministry of Science and Higher Education (grant N N202 3956 33), the Hungarian Scientific Research Fund (OTKA 60506), the Virtual Institute VI-146 of Helmholtz Gemeinschaft, Germany, Korea Research Foundation (KRF-2008-313-C00200), the Federal Agency of Education of the Ministry of Education and Science of the Russian Federation (grant RNP 2.2.2.2.1547) and the Russian Foundation for Basic Research (grant 08-02-00018), the Ministry of Education, Culture, Sports, Science and Technology, Japan, Grant-in-Aid for Scientific Research (18071005,19034011,19740162), Swiss Nationalfonds Foundation 200020-117913/1 and ETH Research Grant TH-01 07-3.
\end{acknowledgments}

\end{document}